\documentclass[pra,showpacs,twocolumn,superscriptaddress]{revtex4}

\newcommand{\ket}[1]{|{#1}\rangle}
\newcommand{\bra}[1]{\langle{#1}|}
\newcommand{\braket}[2]{\langle {#1}|{#2}\rangle}

\newcommand{\be}{\begin{equation}}
\newcommand{\ee}{\end{equation}}

\usepackage{amsfonts}
\usepackage{amsmath}
\usepackage{graphicx}
\usepackage{amssymb}
\usepackage{amsmath}
\usepackage{amssymb}
\usepackage{graphicx}
\usepackage{lscape}
\usepackage{color}
\usepackage{epstopdf}

\begin{document}
\title{Anomalous frequency shifts in a one-dimensional trapped Bose gas}

\author{M. Valiente}
\affiliation{Institute for Advanced Study, Tsinghua University, Beijing 100084, China}
\author{V. Pastukhov}
\affiliation{Department for Theoretical Physics, Ivan Franko National University of Lviv, 12 Drahomanov Street, Lviv-5, 79005, Ukraine}

\begin{abstract}
We consider a system of interacting bosons in one dimension at a two-body resonance. This system, which is weakly interacting, is known to give rise to effective three-particle interactions, whose dynamics is similar to that of a two-dimensional Bose gas with two-body interactions, and exhibits an identical scale anomaly. We consider the experimentally relevant scenario of a harmonically trapped system. We solve the three-body problem exactly and evaluate the shifts in the frequency of the lowest compressional mode with respect to the dipole mode, and find that the effect of the anomaly is to increase the mode's frequency. We also consider the weak-coupling regime of the trapped many-boson problem and find, within the local density approximation, that the frequency of the lowest compressional mode is also shifted upwards in this limit. Moreover, the anomalous frequency shifts are enhanced by the higher particle number to values that should be observable experimentally.
\end{abstract}
\pacs{
}
\maketitle
\section{Introduction}
Systems of trapped ultracold atoms \cite{ReviewBloch} provide a versatile platform for the observation of intriguing few- and many-body phenomena. After the observation of the first Bose-Einstein condensates with alkali atoms over two decades ago \cite{Cornell,Ketterle}, much progress has been made, and both bosonic and fermionic ultracold atomic gases can nowadays be manipulated and controlled with unprecedented accuracy. This includes the tunability of their effective interaction strengths via magnetic \cite{FeshbachReview} and even orbital \cite{ZhangOrbital} Feshbach resonances, or external confinement \cite{Olshanii,NaegerlCIR} and their effective dimensional reduction to one and two dimensions \cite{Kinoshita,Greiner,HallerLuttinger,Krueger}. Among the most interesting phenomena that have been observed using ultracold atoms are the observation \cite{Kraemer,Knoop,Zaccanti} of Efimov states \cite{Efimov}, whose low-energy, model-independent description requires the introduction of three-body forces \cite{Bedaque1,Bedaque2}, the recent observation of the effects of a quantum anomaly in a two-dimensional Fermi gas \cite{Selim2D,Vale2D}, the realisation of an antiferromagnetic few-spin Heisenberg chain \cite{SelimHeisenberg} in the strongly-interacting limit \cite{Volosniev}, and the observation of quantum droplets stabilised by quantum fluctuations \cite{TarruellDroplets,Semeghini}.

Among those systems addressable within the capabilities of current ultracold atomic experiments, reduced one-dimensional systems are especially appealing for a number of reasons. Firstly, there exists a variety of models that are exactly solvable by means of the Bethe ansatz \cite{CazalillaReview}. Among these, the Lieb-Liniger model \cite{LiebLiniger}, consisting of many bosons interacting via zero-range potentials, is of particular relevance to ultracold atoms, since it faithfully describes a variety of interaction and temperature regimes, as has been shown experimentally time and time again \cite{HallerLuttinger,Krueger,Naegerl,Pan}. Secondly, quantum fluctuations are very relevant in one-dimensional systems and lead to the breakdown of Fermi liquid theory for fermions, which are in this case described by the Luttinger liquid theory \cite{HaldaneLuttinger} at low energies. Moreover, the distinction between bosons and fermions is rather blurred in one dimension, with Luttinger liquid theory describing bosons as well, while in the hard-core \cite{Girardeau} and even in certain soft-core \cite{CheonShigehara} limits, fermions and bosons are related via duality transformations -- Bose-Fermi mapping theorems.

When we consider bosons in one dimension, the Lieb-Liniger model is to be regarded as an effective field theory at low energies \cite{EFT,EFTValienteZinner}. In the weakly repulsive case, it describes systems with (negative) scattering lengths that are much larger in magnitude than the interaction's effective range \cite{Adhikari1D}. Effective range effects come into play as a next-to-leading order effect in the ground state. This effect, however, identically vanishes if the scattering length diverges, rendering the many-body system non-interacting to all orders in the two-body effective interactions \cite{ValienteThreeBody}. Typically, one assumes that the non-interacting picture is accurate even in the many-particle limit \cite{Olshanii,Menotti}, which has been shown to be correct within experimental uncertainty in Ref.~\cite{Naegerl} that, however, focused entirely on the opposite, strongly-interacting limit. Interestingly, even in the ground state, next-to-leading order effects do contribute to the physics of the few- and many-boson problems, and these come in the form of three-body contact interactions \cite{ValienteThreeBody,Pricoupenko1,Pricoupenko2,Guijarro,Nishida1,Nishida2}. When the effective three-body forces are attractive, multiparticle bound states may be formed \cite{Pricoupenko1,Pricoupenko2,Nishida1,Nishida2,Drut1,Drut2}, while for three-body repulsion \cite{ValienteThreeBody,Guijarro} a quantum anomaly in the form of logarithmic corrections is found for three or more particles \cite{ValienteThreeBody,Drut1,Drut2,Pricoupenko3}. References \cite{Drut1} and \cite{Drut2} focused on the anomaly within this context. Remarkably, this is completely equivalent to the anomaly found with two-body interactions in two dimensions at low energies \cite{Selim2D,Vale2D,Randeria,Hofmann,Beane2D,LianyiHe}. This can be understood by observing the kinematic equivalence -- identical Schr{\"o}dinger equations -- between three particles in one dimension with three-body contact interactions and two particles in two dimensions with two-body contact interactions.

With all the above theoretical works and the great recent interest in effective three-body interactions in one dimension, it only remains to be seen whether the effects of such forces can be observed experimentally. Since most experiments feature weak harmonic confinement along the effective dimension of the system, with the notable exception of box-like traps \cite{Hadzibabic}, measuring the excitation frequency shifts in the lowest compressional mode is arguably the most suitable way of probing the effects of the three-body interactions. Before that, however, it is necessary to see whether these shifts may be sizeable even in the weak coupling limit. This the goal of the present work.

Here, we begin by reviewing the regularisation and renormalisation of the effective three-body interaction, both in the position and momentum representations, which we use to solve the trapped three-body problem analytically. With the exact solution at hand, we find the excitation frequency of the lowest compressional mode, which is shifted due to the quantum anomaly, and compare the exact results with the sum rule approach of Ref.~\cite{Menotti}, simplified by the generalised virial theorem, which we also derive. We then study the weak-coupling limit of the trapped many-body problem, within the mean-field \cite{Pastukhov} and local density approximations \cite{Dalfovo}, and extract the anomalous frequency shifts in this limit. 

\section{Hamiltonian of the system}
We consider one-dimensional non-relativistic identical bosons with mass $m$ interacting via pairwise potentials. The system is trapped in a harmonic well with frequency $\omega$. We assume that the two-body scattering length $a_2$ is negative and much larger in absolute value than all other length scales in the system. In this way, two particles at low energies become effectively non-interacting, since the Lieb-Liniger interaction strength $g=-2\hbar^2/ma_2$, associated with a two-body effective potential $g\delta(x_1-x_2)$ \cite{LiebLiniger}, vanishes identically. Since two-body interactions are in reality not pointlike, residual interactions remain in the system at either higher energies or for more than two particles. For identical bosons, the lowest-order effects of the interaction when the boson-boson scattering length diverges are due to an effective three-body force \cite{ValienteThreeBody}. Its bare (unrenormalised) form is given by
\begin{equation}
V_3^{\mathrm{LO}}(x_1,x_2,x_3) = g_3 \delta(x_1-x_2)\delta(x_2-x_3).\label{LO}
\end{equation}
The total Hamiltonian with the above interaction is therefore given by
\begin{equation}
  H=H_0+\sum_iV_{\mathrm{trap}}(x_i)+\sum_{i<j<l} V_3^{\mathrm{LO}}(x_i,x_j,x_l),\label{Hamiltonian}
\end{equation}
where
\begin{align}
  &H_0 = -\frac{\hbar^2}{2m}\sum_{i=1}^N \partial_{x_i}^2,\label{H0}\\
  &V_{\mathrm{trap}}(x_i) = \frac{1}{2}m\omega^2x_i^2.
\end{align}

\section{Three-body problem}
Here, we study the three-boson problem with effective three-body forces given by Eq.~(\ref{LO}). We begin by reviewing the three-body problem in free space, whose solution via regularisation-renormalisation can be worked out analytically both in the momentum \cite{Nishida1,Nishida2,Drut1,Drut2,ValienteThreeBody} and position \cite{Pricoupenko2,Guijarro} representations, with an emphasis on the repulsive side of the three-body interaction. We then solve the three-body problem in a harmonic trap analytically, taking advantage of the kinematic equivalence between three bosons in 1D with three-body interactions and two bosons in 2D with two-body interactions \cite{Beane2D}, at low energies \cite{Drut1}.

\subsection{Renormalisation of the bare interaction}
The bare lowest-order interaction, $V_3^{\mathrm{LO}}$, in Eq.~(\ref{LO}) is too singular in one dimension and requires regularisation-renormalisation. 

Since the trap introduces no new singularities, it is sufficient to renormalise the interacting theory in free space. We solve the three-body scattering problem first in momentum space by using the Lippmann-Schwinger equation for the transition matrix (T-matrix) $T(z)$, which reads
\begin{equation}
T(z) = V_3^{\mathrm{LO}} + V_3^{\mathrm{LO}}G_0(z)T(z), \label{LSE}
\end{equation}
where $G_0(z)=(z-H_0)^{-1}$ is the non-interacting Green's function. Since the three-body interaction conserves total momentum $K=k_1+k_2+k_3$, and the system is Galilean relativistic, we rid ourselves of the centre of mass and set its momentum $K=0$. In this frame, the Lippmann-Schwinger equation (\ref{LSE}) becomes two-dimensional, and we find
\begin{equation}
\langle k_1' k_2' ||T(z)|| k_1 k_2 \rangle = g_3+\int \frac{\mathrm{d}q_1\mathrm{d}q_2}{(2\pi)^2} \frac{g_3\langle q_1 q_2 ||T(z)|| k_1 k_2 \rangle}{z-\frac{\hbar^2}{m}(q_1^2+q_2^2+q_1q_2)},\label{LSE2}
\end{equation}
where we have defined the reduced T-matrix via
\begin{equation}
  \bra{k_1'k_2'k_3'}T(z)\ket{k_1k_2k_3}=2\pi \delta(K-K') \langle k_1' k_2' ||T(z)|| k_1 k_2 \rangle,
\end{equation}
and where we have set $K=k_1+k_2+k_3=0$. Obviously, we have chosen plane wave normalisation as $\braket{k'}{k}=2\pi \delta(k-k')$. Since the Fourier transform of the interaction (\ref{LO}) is a constant, we find that $\langle k_1' k_2' ||T(z)|| k_1 k_2 \rangle \equiv t_3(z)$ is a constant (i.e. independent of momentum states and only energy-dependent), and Eq.~(\ref{LSE2}) is readily solved as
\begin{equation}
  t_3(z) = \frac{1}{g_3^{-1}-\mathcal{I}(z)},\label{t3}
\end{equation}
where we have defined
\begin{equation}
\mathcal{I}(z) = \int \frac{\mathrm{d}q_1\mathrm{d}q_2}{(2\pi)^2} \frac{1}{z-\frac{\hbar^2}{m}(q_1^2+q_2^2+q_1q_2)}.\label{Iz}
\end{equation}
Clearly, the integral $\mathcal{I}(z)$ in Eq.~(\ref{Iz}) is ultraviolet (UV) divergent. We must therefore regularise it in the UV in preparation for its renormalisation. To do this, we first change variables in Eq.~(\ref{Iz}) to Jacobi coordinates. We define $q_x=(q_1-q_2)/\sqrt{2}$ and $q_y=\sqrt{2/3}\left[q_3-(q_1+q_2)/2\right] = -\sqrt{3/2}(q_1+q_2)$, where in the last equality we have used $q_3=-q_1-q_2$ ($K=0$). We further transform the integral to polar coordinates via $q_x=q\cos\phi$ and $q_y=q\sin\phi$, and set a hard cut-off $\Lambda$ in the hyperradial momentum integral, obtaining
\begin{equation}
  \mathcal{I}(z) = \frac{1}{\sqrt{3}}\int_0^{\Lambda}\frac{\mathrm{d}q}{2\pi} \frac{q}{z-\frac{\hbar^2}{2m}q^2}=\frac{m}{2\pi\sqrt{3}\hbar^2}\left[\log\left(\frac{k^2}{\Lambda^2}\right)-i\pi\right],\label{IzSolved}
\end{equation}
where we have used $z=E+i0^+$ and where $k^2=2mE/\hbar^2$. Using Eqs.~(\ref{t3}) and (\ref{IzSolved}), we find
\begin{equation}
  t_3(z) = \frac{1}{g_3^{-1}-\frac{m}{2\pi\sqrt{3}\hbar^2}\left[\log\left(\frac{k^2}{\Lambda^2}\right)-i\pi\right]}.\label{t3Solved}
\end{equation}
Since we shall consider repulsive three-body interactions, we set a UV momentum scale, $Q_*$, such that the Landau pole of the T-matrix occurs at energy $E=-\hbar^2Q_*^2/2m$, which would correspond to the trimer's energy in the case of attractive interactions \cite{Pricoupenko1}. We note that the effective theory is valid at energies much lower than $\hbar^2Q_*^2/2m$, and therefore $Q_*^{-1}$ sets the length scale under which the theory ceases to be physically correct. The denominator in Eq.~(\ref{t3Solved}) vanishes if
\begin{equation}
\frac{1}{g_3} = \frac{m}{\pi \sqrt{3} \hbar^2}\log\left(\frac{Q_*}{\Lambda}  \right),\label{g3renormalization}
\end{equation}
which renormalises the T-matrix, rendering it finite. It takes the form
\begin{equation}
t_3(z) = \frac{2\pi\sqrt{3}\frac{\hbar^2}{m}} {\log\left(\frac{Q_*^2}{k^2}\right)+i\pi}.\label{t3Full}
\end{equation}
As is clearly observed in Eq.~(\ref{t3Full}) above, scale invariance at the classical level is broken by quantum renormalisation effects, i.e. the T-matrix exhibits a quantum anomaly \cite{Drut1}.

We now investigate the problem in the position representation, which is necessary for the exact solution of the trapped three-body problem. Here, instead of regularising the problem first, we may use the non-interacting Hamiltonian (\ref{H0}) supplemented with a short-range boundary condition. The simplest way to study this is to investigate the Landau pole or the trimer. First, we perform a change of variables to Jacobi coordinates. We eliminate the centre of mass coordinate $X=(x_1+x_2+x_3)/3$, and set the total momentum to zero. Then we define, as in the momentum representation, $x=(x_1-x_2)/\sqrt{2}$ and $y=\sqrt{2/3}[x_3-(x_1+x_2)/2]$, and go to polar coordinates via $x=r\cos\theta$ and $y=r\sin\theta$. For negative energies, the Schr{\"o}dinger equation for the hyperradial wave function $R(r)$ reads
\begin{equation}
R''(r) + \frac{1}{r}R'(r) -\frac{2m|E|}{\hbar^2}R(r)=0
\end{equation}
The unnormalised singular solution at negative energies is simply given by
\begin{equation}
  R(r) = K_0(2m|E|r/\hbar^2),
\end{equation}
where $K_0$ is the zero-th order modified Bessel function of the second kind \cite{Abramowitz}. We set now the location of the Landau pole to $E=-\hbar^2Q_*^2/2m$, and using the short distance expansion of $K_0$, we obtain the desired boundary condition
\begin{equation}
  R(r) = - \log\left(\frac{Q_*e^{\gamma}r}{2}\right)+O(\log(Q_*r)(Q_*r)^2),\label{Boundary}
\end{equation}
where $\gamma$ is Euler's gamma constant. We also link the value of the UV scale to the three-body scattering length $a_3$, via \cite{ValienteThreeBody,Pastukhov}
\begin{equation}
  (Q_*a_3)^2 = 8e^{-2\gamma},\label{Qtoa3}
\end{equation}
since we will use the latter, instead of $Q_*$, in the many-body problem.

\subsection{Exact solution of the three-body problem in a harmonic trap}
We now present the solution to the three-body problem in the presence of harmonic confinement. This is equivalent \cite{Busch} to the problem of a single particle of mass $m$ in two dimensions in a harmonic trap with frequency $\omega$ with a contact interaction, whose short-range boundary condition is given by Eq.~(\ref{Boundary}), as we shall see shortly. In Jacobi coordinates $(X,x,y)$, the non-interacting Hamiltonian reads
\begin{equation}
  H_0 = H_{\mathrm{CM}} + H_{\mathrm{r}},
\end{equation}
with
\begin{align}
  &H_{\mathrm{CM}}= -\frac{\hbar^2}{6m}\partial_X^2 + \frac{3}{2}m\omega^2X^2,\label{CMHamiltonian}\\
  &H_{\mathrm{r}}=-\frac{\hbar^2}{2m} (\partial_x^2+\partial_y^2) + \frac{1}{2}m\omega^2(x^2+y^2).
\end{align}
For the sake of completeness, the bare two-body interaction is transformed as
\begin{equation}
  V_3^{\mathrm{LO}} = g_3\delta(\sqrt{2}x)\delta(\sqrt{3/2}y) = \frac{g_3}{\sqrt{3}}\delta(x)\delta(y).
\end{equation}
The centre of mass motion is governed by Hamiltonian (\ref{CMHamiltonian}), which corresponds to a particle of mass $3m$ with frequency $\omega$, and is readily solved. For the relative motion, together with the boundary condition (\ref{Boundary}), we go to polar coordinates and define
\begin{equation}
R(r) = \exp\left({-\frac{r^2}{2a_{\parallel}^2}}\right)F(r),\label{Rofr}
\end{equation}
where $a_{\parallel}=\sqrt{\hbar/m\omega}$ is the harmonic length. Defining $s=(r/a_{\parallel})^2$ and $F(r)=u(s)$, the Schr{\"o}dinger equation for angular momentum $m_z=0$ reads
\begin{equation}
  su''(s) + (1-s)u'(s) +\frac{\mathcal{E}-1}{2}u(s) = 0,
\end{equation}
where $\mathcal{E}=E/\hbar \omega$ is the dimensionless form of the (relative) energy eigenvalues. The singular solution that leaves $R(r)$, Eq.~(\ref{Rofr}), normalisable is given by the following confluent hypergeometric function $U$ \cite{Abramowitz}
\begin{equation}
u(s) = U\left(\frac{1-\mathcal{E}}{2},1,s\right).\label{uSolution}
\end{equation}
At short distances ($s\ll 1$), the confluent hypergeometric function behaves as
\begin{equation}
  U(a,1,s) \propto -\log s - \psi(a) - 2\gamma + O(s\log s),\label{Uprop}
\end{equation}
where $\psi$ is the digamma function. Comparing the boundary condition (\ref{Boundary}) with the short-range form of $u(s)$, Eqs.~(\ref{uSolution}) and (\ref{Uprop}), we obtain the eigenvalue equation
\begin{equation}
  \psi\left(\frac{1-\mathcal{E}}{2}\right) = \log\left[\left(\frac{Q_*a_{\parallel}}{2} \right)^2 \right].\label{EigenvalueEquation}
\end{equation}

\section{Tan's contact and the Virial theorem}\label{Section:Virial}
We here derive the virial theorem for our system, which will be useful in the next section. This involves the so-called Tan's contact $C_3$ \cite{Tan1,Tan2,Tan3,Drut1,Drut2,Pastukhov}, relating the large-momentum tail of the momentum distribution, or the short-distance correlation functions, to a number of physical quantities. In our case, the virial theorem reads
\begin{equation}
  E = 2\langle V_{\mathrm{trap}} \rangle -\frac{\hbar^2}{4\pi m} \int \mathrm{d}x C_3(x),\label{Virial}
\end{equation}
where $E$ is the energy and $C_3(x)$ is the local three-body contact, given by
\begin{equation}
  C_3(x) = \frac{1}{3\sqrt{3}} \left(\frac{mg_3}{\hbar^2}\right)^2 \langle \left[\phi^{\dagger}(x) \right]^3\left[\phi(x)\right]^3\rangle.
\end{equation}
Above, $\phi^{\dagger}$ and $\phi$ are bosonic creation and annihilation operators in the position representation, while $g_3$ is the bare coupling constant given by Eq.~(\ref{g3renormalization}).

To derive Eq.~(\ref{Virial}), we use a method analogous to Ref.~\cite{Werner}, although this can be derived using Tan's original method particularised to two dimensions\cite{ValienteZinnerMolmer1,ValienteZinnerMolmer2}. We begin by calling $\Psi(x_1,\ldots,x_N)$ the normalised $N$-boson ground state of Hamiltonian (\ref{Hamiltonian}) with energy $E$. We rescale the coordinates of the particles as $x_i\to \lambda x_i$ ($i=1,\ldots,N$) in $\Psi$, and evaluate how the expectation value of the energy, $E(\lambda)$, in the rescaled state $\Psi_{\lambda}$, changes due to this transformation. The rescaled state must be multiplied by a constant in order to remain normalised, as
\begin{equation}
  \Psi_{\lambda}(x_1,\ldots,x_N) = \lambda^{N/2} \Psi(\lambda x_1,\ldots,\lambda x_N).
\end{equation}
The rescaled wave function $\Psi_{\lambda}$ does not satisfy the same short-distance boundary condition, Eq.~(\ref{Boundary}), as the eigenstate $\Psi=\Psi_{1}$. In order to fix this, we must rescale the UV momentum scale $Q_*$ as $Q_*^{-1}\to \lambda Q_*^{-1}$. The expectation value of the energy in this state reads
\begin{equation}
  E(\lambda) = \lambda^2\langle H_0 \rangle_1 + \lambda^{-2}\langle V_{\mathrm{trap}} \rangle_1 + \lambda^2\langle V_3^{\mathrm{LO}} \rangle_{\lambda},\label{Elambda}
\end{equation}
where $\langle \cdot \rangle_{\lambda}$ stands for expectation value in state $\Psi_{\lambda}$. Using the fact that $E(\lambda)$ is minimized for $\lambda=1$, corresponding to the true ground state of the system, we now differentiate $E(\lambda)$ in Eq.~(\ref{Elambda}) with respect to $\lambda$ and obtain
\begin{align}
&2\langle H_0 \rangle -2 \langle V_{\mathrm{trap}} \rangle +2\langle V_3^{\mathrm{LO}} \rangle \nonumber \\
  &+\frac{\partial_{\lambda}g_3(\lambda\Lambda)\left.\right|_{\lambda=1}}{3!}\int \mathrm{d}x \langle \left[\phi^{\dagger}(x)\right]^3 \left[\phi(x)\right]^3\rangle=0,\label{someeq}
\end{align}
where we have set once more $\langle \cdot \rangle = \langle \cdot \rangle_1$, and where we have used the relation between first and second-quantised operators
\begin{equation}
  \sum_{i<j<l}\langle \delta(x_i-x_j)\delta(x_i-x_l)\rangle = \frac{1}{3!} \int  \mathrm{d}x \langle \left[\phi^{\dagger}(x)\right]^3 \left[\phi(x)\right]^3\rangle.
\end{equation}
Substituting Eq.~(\ref{g3renormalization}) into (\ref{someeq}) and then setting $\lambda=1$ in Eq.~(\ref{Elambda}) (note that $E(1)=E$), we finally obtain the virial theorem, Eq.~(\ref{Virial}).

Since we shall also need the adiabatic relation, which was derived in Ref.~\cite{Pastukhov}, we simply quote it here, particularised to the trapped case
\begin{equation}
  -\frac{\partial E}{\partial \log Q_*} = \frac{\hbar^2}{2\pi m} \int \mathrm{d}x C_3(x).\label{Adiabatic}
\end{equation}

\section{Excitations in the few-body limit}
With the exhaustive analysis of the three-body problem performed above, we are ready to calculate the excitation frequencies in the three-body sector. This will give us the order of magnitude of the shifts in frequencies due to the three-body interaction in the few-body limit, where the local density approximation (LDA) is not valid. This regime is experimentally relevant, as quasi-one-dimensional geometries consisting of arrays of tubes with few ($N\sim 8-11$) particles can be routinely prepared for at least a decade now \cite{Naegerl}.

For three particles, the lowest mode, in the non-interacting picture, that is affected by interactions corresponds to $E=3\hbar \omega$. This is called the lowest compressional mode \cite{Menotti}, with a non-interacting excitation frequency $\omega_C^{(0)}=2\omega$. The dipole mode \cite{Menotti} is not affected by the interactions and has energy $E=2\hbar \omega$. Here, we shall explore the shifts in the ratio $R=\omega_C^2/\omega_D^2$ between the square frequency of the lowest compressional and the dipole modes due to the effective three-body interactions. The shift is defined as $\Delta = R-\left(\frac{\omega_C^{(0)}}{\omega_D}\right)^2 = R-4$.

Before discussing the exact solution to Eq.~(\ref{EigenvalueEquation}), we study its weak-coupling limit. To this end, we use the Laurent expansion of the digamma function $\psi(a)$ near its poles $a=0$ and $a=-1$ \cite{Abramowitz}. These read
\begin{align}
  \psi(a) &= -\frac{1}{a} - \gamma + \frac{\pi^2}{6}a+O(a^2),\hspace{0.2cm} a\to 0,\\
  \psi(a) &= -\frac{1}{a+1}+1-\gamma +\left(1+\frac{\pi^2}{6}\right)(a+1)\nonumber \\
          &+ O((a+1)^2),\hspace{0.2cm} a\to 1. 
\end{align}
Using $a=(1-\mathcal{E})/2$ in the above relations, together with the eigenvalue equation (\ref{EigenvalueEquation}), we find for the ground and second (first for bosons) excited states
\begin{align}
\mathcal{E}_0 &= 1+g_{\omega}-\frac{\gamma}{2}g_{\omega}^2+O(g_{\omega}^3),\label{E0Weak}\\
\mathcal{E}_2 &= 3+g_{\omega}+\frac{1-\gamma}{2}g_{\omega}^2 + O(g_{\omega}^3).\label{E2Weak}
\end{align}
Above, we have defined the dimensionless coupling constant $g_{\omega}$ as
\begin{equation}
g_{\omega} = \frac{1}{\log\left(\frac{Q_*a_{\parallel}}{2} \right)}.\label{gomega}
\end{equation}
We now see that, to lowest order in the interaction coupling constant (\ref{gomega}), the excitation frequency of the lowest compressional mode is given by
\begin{equation}
\frac{\omega_C}{\omega} = \mathcal{E}_2-\mathcal{E}_0 \approx 2+\frac{1}{2}g_{\omega}^2. 
\end{equation}
As for the dipole mode, which is not affected by interactions, we simply have $\omega_D=\omega$. Therefore, the weak-coupling limit of the shift $\Delta$ in the ratio $R$, with respect to the non-interacting case, is given by
\begin{equation}
  \Delta = 2g_{\omega}^2 + O(g_{\omega}^3).\label{DeltaWeak}
\end{equation}

We also use another route to obtain Eq.~(\ref{DeltaWeak}) by combining sum rules with the theory of Tan's contact (see Section~\ref{Section:Virial}). An upper bound to the ratio $R$ is obtained via the expression (see Ref.~\cite{Menotti})
\begin{equation}
  \omega_C^2 = -2\frac{\langle x^2 \rangle}{\frac{\mathrm{d}\langle x^2 \rangle}{\mathrm{d}\omega^2}}.\label{Menotti}
\end{equation}
Using the Hellmann-Feynman theorem, we can relate the expectation value of $x^2$ in the ground state with the derivative of the ground state energy $E_0$ (with respect to the centre-of-mass zero-point energy) with respect to $\omega^2$, as
\begin{equation}
  \frac{\mathrm{d}E_0}{\mathrm{d}\omega^2} = \frac{3}{2}m\langle x^2 \rangle.\label{x2-1}
\end{equation}
The Virial theorem, Eq.~(\ref{Virial}), combined with the adiabatic relation \cite{Pastukhov}, Eq.~(\ref{Adiabatic}), gives us another relation between the energy and $\langle x^2 \rangle$, as
\begin{equation}
\langle x^2 \rangle = \frac{1}{3m\omega^2} \left[E_0 - \frac{1}{2} \frac{\mathrm{d}E_0}{\mathrm{d}(\log Q_*)}\right].\label{x2-2}
\end{equation}
Simple algebraic manipulations yield
\begin{equation}
R=4\frac{E_0(g_{\omega})+\frac{1}{2}g_{\omega}^2E_0'(g_{\omega})}{E_0-\frac{1}{2}g_{\omega}^3E_0'(g_{\omega})-\frac{1}{4}g_{\omega}^4E_0''(g_{\omega})}\label{REnergy}
\end{equation}  
Using it to lowest order in $g_{\omega}$, we find that
\begin{equation}
  \Delta \approx 2\frac{g_{\omega}^2}{E_0}\frac{\mathrm{d}E_0}{\mathrm{d}g_{\omega}}\left.\right|_{g_{\omega=0}}\label{DeltaWeak-2}
\end{equation}
Inserting Eq.~(\ref{E0Weak}) into Eq.~(\ref{DeltaWeak-2}), we find
\begin{equation}
  \Delta = 2g_{\omega}^2 + O(g_{\omega}^3),
\end{equation}
in perfect agreement with the result obtained using the excitation energy directly, Eq.~(\ref{DeltaWeak}).

\begin{figure}[t]
\includegraphics[width=0.5\textwidth]{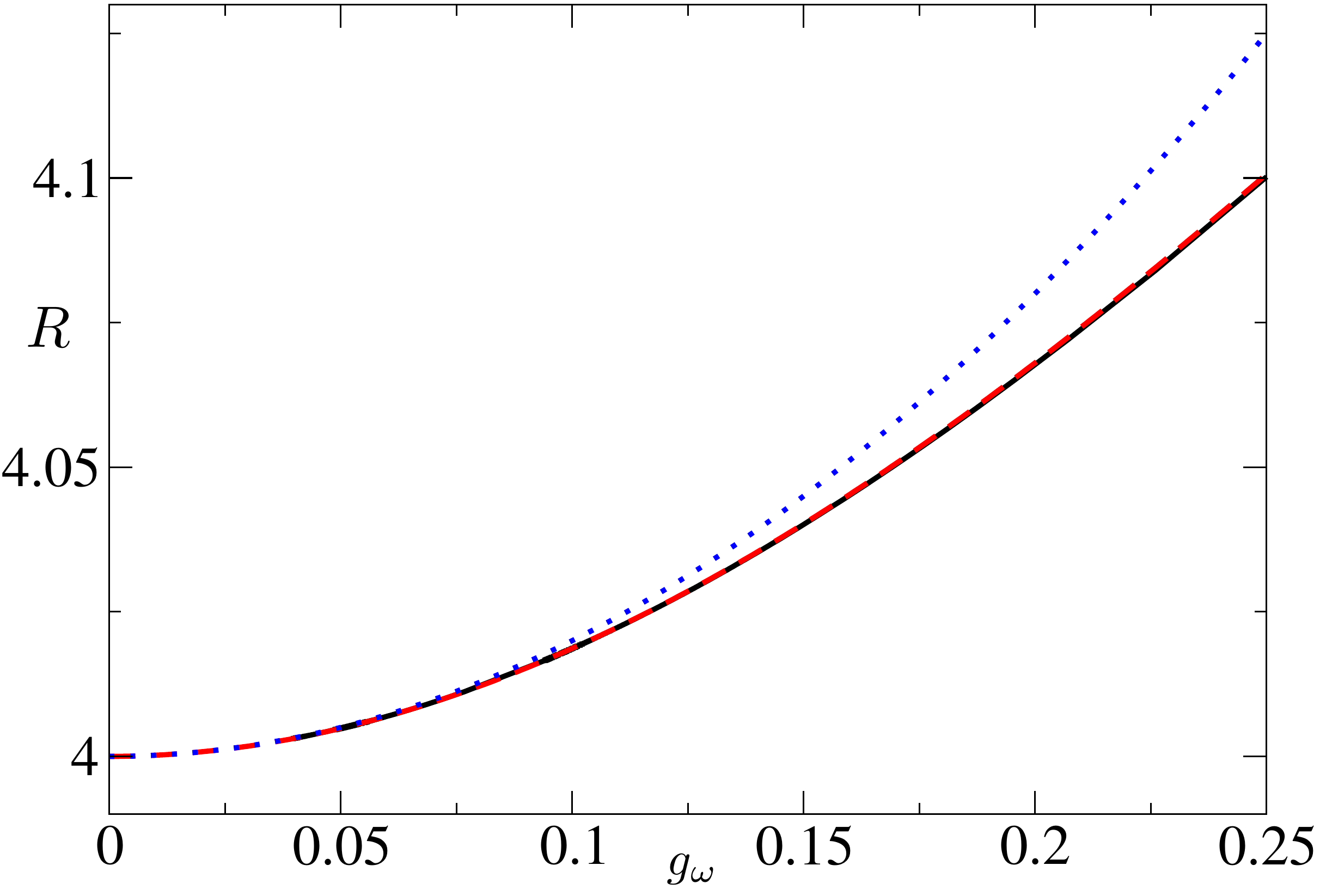}
\caption{Ratio $R=(\omega_C/\omega_D)^2$ between the frequencies of the lowest compressional mode and the dipole mode in the three-body problem. Black solid, red dashed and blue dotted lines correspond, respectively, to $R$ calculated via the excitation spectrum, Eq.~(\ref{EigenvalueEquation}), the sum rule approach, Eq.~(\ref{REnergy}), and the weak-coupling limit, Eq.~(\ref{DeltaWeak}).}
\label{fig:R-figure}
\end{figure}

In Fig.~\ref{fig:R-figure} , we show the value of the ratio $R=(\omega_C/\omega)^2$ as a function of the coupling constant $g_{\omega}$, Eq.~(\ref{gomega}), calculated using the exact excitation spectrum from the solutions of Eq.~(\ref{EigenvalueEquation}), the sum rule approach, Eq.~(\ref{REnergy}), and the weak-coupling limit, Eqs.~(\ref{DeltaWeak}) and (\ref{DeltaWeak-2}). We see that, in the very weak coupling regime, the shift $\Delta = R-4$ is approximately quadratic in the coupling constant. Moreover, the agreement between the results using the excitation spectrum and the sum rules is excellent. We clearly see that the frequency $\omega_C$ is shifted upwards with respect to its non-interacting value $\omega_C^{(0)}=2\omega$. This is to be contrasted with the frequency shift due to repulsive Lieb-Liniger interactions which, for the few-body problem at weak coupling, is downwards \cite{Busch}. Therefore, the observation of a positive energy shift in this limit would be a clear signature of the role of effective three-body forces in 1D Bose systems.

\section{Many-body problem}
We now study the many-particle limit. Since the effective three-body repulsive interactions appear to be weak in all the physical settings studied so far \cite{ValienteThreeBody,Guijarro}, we can safely assume that mean-field theory will be at least qualitatively valid. In Ref.~\cite{Pastukhov}, one of us derived the mean-field, Bogoliubov and beyond mean-field corrections to the ground state energy of the Bose gas with repulsive three-body interactions in the thermodynamic limit. The ground state energy per particle $E/N$ takes the form
\begin{align}
  \frac{E}{N} = &-\frac{\pi\hbar^2n^2}{2\sqrt{3}m\log(na_3)}\left[1-\frac{4\cdot 3^{1/4}}{\sqrt{-\pi\log(na_3)}}\right.\nonumber\\
    &\left.+\frac{\frac{\log[-\log(na_3)]}{2}-C_E}{\log(na_3)}\right] + O(n^2\log^{-2}(na_3)),\label{EoverN}
\end{align}
where $n=N/L$ is the density of the system, and $C_E$ is a numerical constant whose value \cite{Pastukhov} is inconsequential for our purposes. Below, we employ the local density approximation (LDA) to study the large-$N$ limit, in the mean-field approximation of Eq.~(\ref{EoverN}), when the system is placed in a harmonic trap, and calculate the frequency of the lowest compressional mode to this order.

\subsection{Local density approximation}  
Here we deal with the trapped many-body problem within the LDA. We present a fully analytical treatment in the weak-coupling limit, which introduces the quantum anomaly self-consistently. To lowest order in the interaction, the energy per particle $E/N$ of the homogeneous many-body problem at zero temperature is given by (see Eq.~(\ref{EoverN}))
\begin{equation}
  \frac{E}{N} = -\frac{\pi \hbar^2}{2\sqrt{3}m} n^2\left[\frac{1}{\log(na_3)}+O(\log^{-3/2}(na_3))\right].
\end{equation}
The LDA consists of setting the chemical potential $\mu = \mu_{\ell}+V_{\mathrm{trap}}$ \cite{Dalfovo}, where $\mu_{\ell}$ is the local chemical potential, obtained by making the density explicitly dependent on position, i.e. by replacing $n\to n(x)$. We have, to lowest order
\begin{equation}
  \mu = -\frac{\sqrt{3} \pi\hbar^2}{2m}\frac{[n(x)]^2}{\log[n(x)a_3]}+\frac{1}{2}m\omega^2x^2.\label{muLDA1}
\end{equation}
In the weak-coupling limit, the logarithmic term above varies slowly. We can therefore rewrite the exact density $n(x)$ as a mean-field part plus a classical fluctuation
\begin{equation}
  n(x) = n_0 + \frac{\partial_x \phi(x)}{\pi},
\end{equation}
with $n_0$ the average density of the system and $\phi$ the fluctuating field, well known from bosonization \cite{Matveev}, and insert it into the dimensionless coupling constant $-\log^{-1}[n(x)a_3]$ to obtain
\begin{equation}
  -\frac{1}{\log[n(x)a_3]}\approx -\frac{1}{\log(n_0a_3)}\left[1-\frac{\partial_x\phi}{\pi n_0\log(n_0a_3)}\right].\label{FluctuationExpansion}
\end{equation}
The second term in the right hand side of Eq.~(\ref{FluctuationExpansion}) is of higher order than Eq.~(\ref{muLDA1}) and can be dropped in this approximation. Therefore, to lowest order in the three-body interaction, Eq.~(\ref{muLDA1}) is simplified to
\begin{equation}
  \mu = -\frac{\sqrt{3} \pi\hbar^2}{2m}\frac{[n(x)]^2}{\log(n_0a_3)}+\frac{1}{2}m\omega^2x^2,\label{muLDA2}
\end{equation}
with the average density $n_0$ to be obtained self-consistently via
\begin{equation}
n_0 = \langle n(x) \rangle .\label{selfconsistent}
\end{equation}
For further convenience, we define the interaction coupling constant $G(\omega)$ as
\begin{equation}
  G(\omega) = -\frac{\sqrt{3}\pi}{2}\frac{1}{\log(n_0a_3)},\label{gomegatrap}
\end{equation}
where its dependence on the trap's frequency $\omega$ has been made explicit as it is relevant for the calculation of the frequency of the lowest compressional mode. As usual, the chemical potential can be eliminated from Eq.~(\ref{muLDA2}) by setting the density to zero beyond a certain distance $\Gamma$ -- the Thomas-Fermi radius -- and we have $\mu = m\omega^2\Gamma^2/2$. Finally, from Eq.~(\ref{muLDA2}) we have the density profile
\begin{equation}
  n(x) = \frac{1}{\sqrt{2G(\omega)}}\frac{\Gamma}{a_{\parallel}^2}\sqrt{1-\left(\frac{x}{\Gamma}\right)^2}.
\end{equation}
Above, we see that the density profile coincides with that of a scale invariant system, such as the free Fermi gas or the Tonks-Girardeau gas. However, the important difference is the dependence of the coupling constant $G=G(\omega)$ on the frequency of the trap via the self-consistency condition (\ref{selfconsistent}). This fact will introduce an anomalous deviation of the excitation frequency with respect to the scale invariant case (which has $R=4$). We use now the particle number normalisation condition
\begin{equation}
  N=\int_{-\Gamma}^{\Gamma} \mathrm{d}x n(x),
\end{equation}
to eliminate the Thomas-Fermi radius $\Gamma$ (or, equivalently, the chemical potential $\mu$) in favor of $N$. We have
\begin{equation}
  \Gamma^2 = \frac{2\sqrt{2G(\omega)}}{\pi}a_{\parallel}^2 N
\end{equation}
With this, it is simple to extract the squared-mean-radius of the system
\begin{equation}
  \langle x^2 \rangle = \frac{1}{N}\int_{-\Gamma}^{\Gamma}\mathrm{d}x x^2 n(x) = \frac{\sqrt{G(\omega)}N}{\sqrt{2}\pi}a_{\parallel}^2.
\end{equation}
and the mean density $n_0$ from Eq.~(\ref{selfconsistent}),
\begin{equation}
  n_0 =\frac{2^{13/4}}{3\pi^{3/2}}\frac{N^{1/2}}{[G(\omega)]^{1/4}}a_{\parallel}^{-1}.\label{selfconsistency2}
\end{equation}
Inserting Eq.~(\ref{gomegatrap}) into Eq.~(\ref{selfconsistency2}) above, and multiplying both resulting sides by $a_3$, gives us the following equation for the dimensionless parameter $n_0a_3$
\begin{equation}
  n_0a_3 = B\left(\frac{a_3}{a_{\parallel}}\right)\left[-\log(n_0a_3)\right]^{1/4},\label{n0a3Eq}
\end{equation}
with
\begin{equation}
  B=\frac{2^{7/2}N^{1/2}}{3^{9/8}\pi^{7/4}}.
\end{equation}
Eq.~(\ref{n0a3Eq}) is solved by
\begin{equation}
(n_0a_3)^4 = \frac{B^4}{4}\left(\frac{a_3}{a_{\parallel}}\right)^4W\left(\frac{4a_{\parallel}^4}{B^4a_3^4}\right),\label{n0a34}
\end{equation}
where $W$ is Lambert's function \cite{Lambert}. Inserting Eq.~(\ref{n0a34}) into Eq.~(\ref{gomegatrap}), we obtain the following for the coupling constant
\begin{equation}
  G(\omega) = \frac{2\pi\sqrt{3}}{W\left(\frac{4a_{\parallel}^4}{B^4a_3^4}\right)}.
\end{equation}
The ratio $R=(\omega_C/\omega_D)^2$ is readily obtained using Eq.~(\ref{Menotti}), and we have
\begin{equation}
R = 4\left[1-\frac{G(\omega)}{2\pi\sqrt{3}+G(\omega)}\right]^{-1},
\end{equation}
which, to lowest order in $G(\omega)$, takes the form
\begin{equation}
  R= 4 + \frac{2}{\pi\sqrt{3}}G(\omega) + O(G(\omega)^2).\label{RweakMF}
\end{equation}
\begin{figure}[t]
\includegraphics[width=0.5\textwidth]{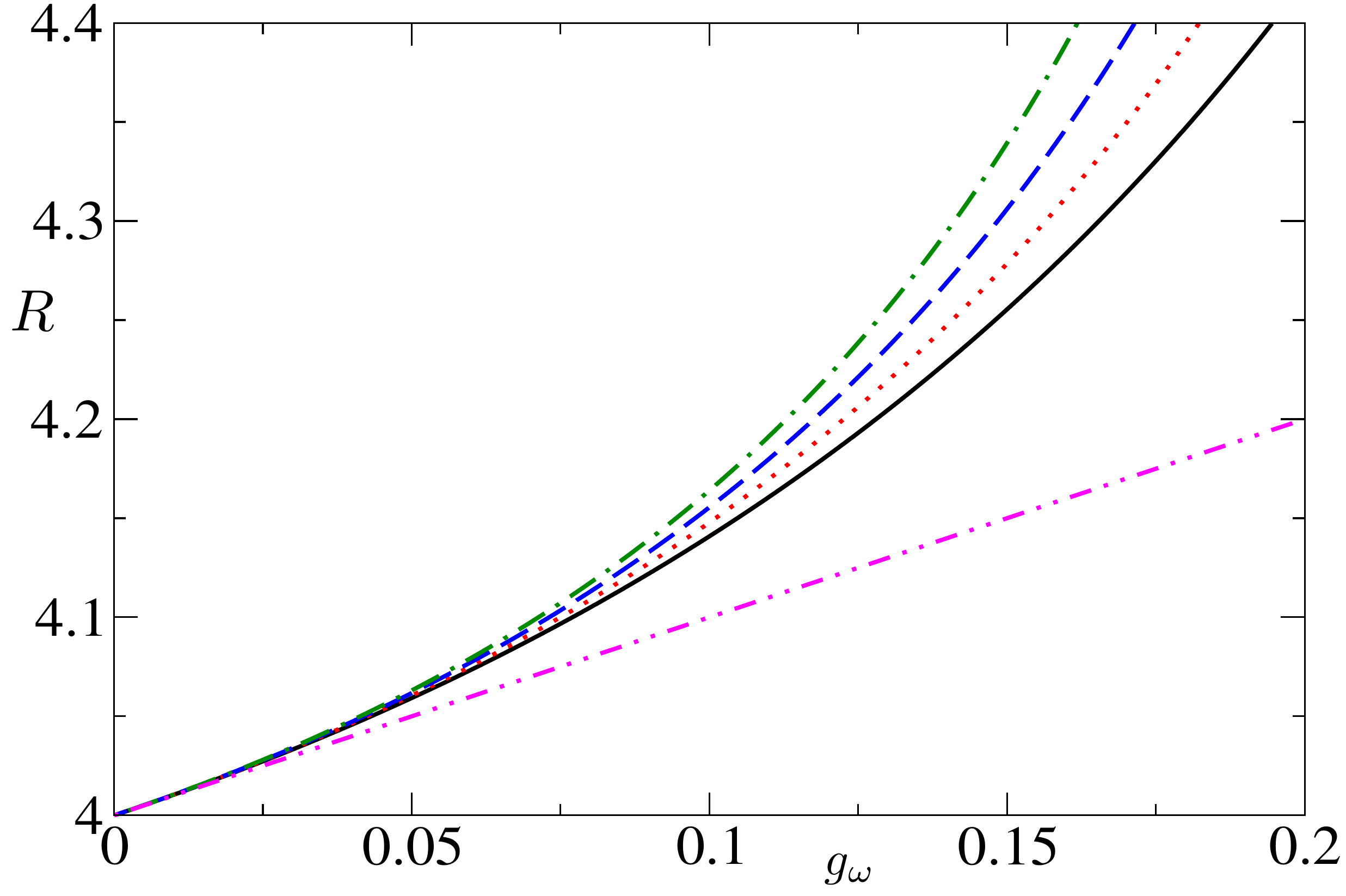}
\caption{Ratio $R=(\omega_C/\omega_D)^2$ between the frequencies of the lowest compressional mode and the dipole mode in the mean-field approximation, as a function of the few-body coupling constant $g_{\omega}$. Black solid, red dotted, blue dashed and green dashed-dotted lines correspond, respectively, to particle numbers $N=250$, 500, 1000 and 2000. The magenta dashed-double dotted line is the extreme weak-coupling result $R\approx 4+g_{\omega}$.}
\label{fig:R-figure-MF}
\end{figure}
The above relation shows that, for repulsive three-body interaction, the role of the anomaly is to shift the frequency of the lowest compressional mode upwards not only in the few-body limit, but also in the mean-field, many-body limit. Moreover, the shift $\Delta = R-4$ is, in this case, of linear order in the coupling constant. This is to be contrasted with the few-body limit, where we showed that the shift, though positive as well, is quadratic in the few-body coupling constant $g_{\omega}$. In the very weak coupling limit, Eq.~(\ref{RweakMF}) becomes independent of the particle number and reads simply $R\approx 4+g_{\omega}$. In Fig.~\ref{fig:R-figure-MF}, we plot the ratio $R$ as a function of the few-body coupling constant, Eq.~(\ref{gomega}), for different particle numbers. As seen there, for very small values of $g_{\omega}$, all curves collapse and they exhibit their $N$-dependence as the coupling becomes stronger. From Fig.~(\ref{fig:R-figure-MF}), it is also clear that larger particle numbers enhance the anomalous frequency shift. For instance, comparing Fig.~\ref{fig:R-figure} with Fig.~\ref{fig:R-figure-MF}, we observe that in order to achieve a shift $\Delta\approx 0.1$ in the three-body problem, we need $g_{\omega}\approx 0.25$, while with 250--2000 particles $g_{\omega}\approx 0.075$ is sufficient. With the experimental values reported for the trapping frequencies in Ref.~\cite{Naegerl}, together with the theoretical value of $a_3$ in the example of Ref.~\cite{ValienteThreeBody}, we may expect a shift of around 3--4\% in the value of $R$ from the scale invariant limit $R=4$. Therefore, if the experimental uncertainties for $R$ are reduced in the weak-coupling limit and, especially, if the atom numbers can be increased from the few-body limit with $N\sim 10$ \cite{Naegerl}, the anomalous shifts should be clearly observable.  

\section{Conclusions}
We have considered one-dimensional bosons with very large two-body scattering lengths trapped in a harmonic well. We have studied the few-body limit by solving the three-body problem exactly, and obtained the shifts, which are always positive, in the frequency of the lowest compressional mode due to the emergent three-body forces among the bosons. We have shown that the sum-rule approach to obtaining the excitation frequencies is in excellent agreement with the exact results. We have also studied the weak-coupling regime of the many body problem in a harmonic trap, and included the effect of the anomaly via a self-consistent version of the local density approximation. We have shown that the frequency shifts are also upwards and largely enhanced by higher particle numbers. Our results open promising avenues to explore three-body forces experimentally using trapped ultracold atoms.

\bibliographystyle{unsrt}

\begin{thebibliography}{99}
\bibitem{ReviewBloch} I. Bloch, J. Dalibard and W. Zwerger, Rev. Mod. Phys. {\bf 80}, 885 (2008).

\bibitem{Cornell} M.~H. Anderson, J.~R. Ensher, M.~R. Matthews, C.~E. Wieman and E.~A. Cornell, Science {\bf 269}, 5221 (1995).

\bibitem{Ketterle} K.~B. Davis, M.~-O. Mewes, M.~R. Andrews, N.~J. van Druten, D.~S. Durfree, D.~M. Kurn and W. Ketterle, Phys. Rev. Lett. {\bf 75}, 3969 (1995).

\bibitem{FeshbachReview} C. Chin, R. Grimm, P. Julienne and E. Tiesinga, Rev. Mod. Phys. {\bf 82}, 1225 (2010).

\bibitem{Olshanii} M. Olshanii, Phys. Rev. Lett. {\bf 81}, 938 (1998).

\bibitem{NaegerlCIR} E.~Haller {\it et al.}, Phys. Rev. Lett. {\bf 104}, 153203 (2010).  

\bibitem{ZhangOrbital} R. Zhang, Y. Cheng, H. Zhai and P. Zhang, Phys. Rev. Lett. {\bf 115}, 135301 (2015).

\bibitem{Kinoshita} T. Kinoshita, T. Wenger and D.~S. Weiss, Science {\bf 305}, 1125 (2004).

\bibitem{Greiner} M. Greiner, I. Bloch, O. Mandel, T.~W. H{\"a}nsch and T. Esslinger, Phys. Rev. Lett. {\bf 87}, 160405 (2001).

\bibitem{HallerLuttinger} E. Haller {\it et al.}, Nature {\bf 466}, 597 (2010).

\bibitem{Krueger} P. Kr{\"u}ger, S. Hofferberth, I.~E. Mazets, I. Lesanovsky and J. Schmiedmayer, Phys. Rev. Lett. {\bf 105}, 265302 (2010).

\bibitem{Kraemer} T. Kraemer {\it et al.}, Nature {\bf 440}, 315 (2006).

\bibitem{Knoop} S. Knoop {\it et al.}, Nature Phys. {\bf 5}, 227 (2009).

\bibitem{Zaccanti} M. Zaccanti {\it et al.}, Nature Phys. {\bf 5}, 586 (2009).

\bibitem{Efimov} V. Efimov, Phys. Lett. B {\bf 33}, 563 (1970).

\bibitem{Bedaque1} P.~F. Bedaque, H.~-W. Hammer and U. van Kolck, Phys. Rev. Lett. {\bf 82}, 463 (1999).

\bibitem{Bedaque2} P.~F. Bedaque, H.~-W. Hammer and U. van Kolck, Nucl. Phys. A {\bf 646}, 444 (1999).

\bibitem{Selim2D} M. Holten, L. Bayha, A.~C. Klein, P.~A. Murthy, P.~M. Preiss and S. Jochim, Phys. Rev. Lett. {\bf 121}, 120401 (2018).

\bibitem{Vale2D} T. Peppler, P. Dyke, M. Zamorano, S. Hoinka and C.~J. Vale, Phys. Rev. Lett. {\bf 121}, 120402 (2018).

\bibitem{SelimHeisenberg} S. Murmann {\it et al.}, Phys. Rev. Lett. {\bf 115}, 215301 (2015).

\bibitem{Volosniev} A.~G. Volosniev, D.~V. Fedorov, A.~S. Jensen, M. Valiente and N.~T. Zinner, Nature Comms. {\bf 5}, 5300 (2014).

\bibitem{TarruellDroplets} C.~R. Cabrera, L. Tanzi, J. Sanz, B. Naylor, P. Thomas and L. Tarruell, Science {\bf 359}, 301 (2018).

\bibitem{Semeghini} G. Semeghini {\it et al.}, Phys. Rev. Lett. {\bf 120}, 235301 (2018).

\bibitem{CazalillaReview} M.~A. Cazalilla, R. Citro, T. Giamarchi, E. Orignac and M. Rigol, Rev. Mod. Phys. {\bf 83}, 1405 (2011).
  
\bibitem{LiebLiniger} E.~H. Lieb and W. Liniger, Phys. Rev. {\bf 130}, 1605 (1963).

\bibitem{Naegerl} E. Haller {\it et al.}, Science {\bf 325}, 1224 (2009).
  
\bibitem{Pan} B. Yang {\it et al.}, Phys. Rev. Lett. {\bf 119}, 165701 (2017).

\bibitem{HaldaneLuttinger} F.~D.~M. Haldane, J. Phys. C: Solid State Phys. {\bf 14}, 2585 (1981).
  
\bibitem{Girardeau} M. Girardeau, J. Math. Phys. {\bf 1}, 516 (1960).

\bibitem{CheonShigehara} T. Cheon and T. Shigehara, Phys. Rev. Lett. {\bf 82}, 2536 (1999).

\bibitem{EFT} E. Epelbaum, H.~-W. Hammer and U.~-G. Mei{\ss}ner, Rev. Mod. Phys. {\bf 81}, 1773 (2009).

\bibitem{EFTValienteZinner} M. Valiente and N.~T. Zinner, Few-body syst. {\bf 56}, 845 (2015).
  
\bibitem{Adhikari1D} V.~E. Barlette, M.~M. Leite and S. Adhikari, Eur. J. Phys. {\bf 21}, 435 (2000).

\bibitem{ValienteThreeBody} M. Valiente, e-print arXiv:1902.01643v1 .

\bibitem{Menotti} C. Menotti and S. Stringari, Phys. Rev. A {\bf 66}, 043610 (2002).

\bibitem{Pricoupenko1} L. Pricoupenko, Phys. Rev. A {\bf 97}, 061604 (2018).

\bibitem{Pricoupenko2} L. Pricoupenko, Phys. Rev. A {\bf 99}, 012711 (2019).

\bibitem{Guijarro} G. Guijarro, A. Pricoupenko, G.~E. Astrakharchik, J. Boronat and D.~S. Petrov, Phys. Rev. A {\bf 97}, 061605 (2018). 

\bibitem{Nishida1} Y. Sekino and Y. Nishida, Phys. Rev. A {\bf 97}, 011602 (2018).

\bibitem{Nishida2} Y. Nishida, Phys. Rev. A {\bf 97}, 061603 (2018).
  
\bibitem{Drut1} J.~E. Drut, J.~R. McKenney, W.~S. Daza, C.~L. Lin and C.~R. Ord{\'o}{\~n}ez, Phys. Rev. Lett. {\bf 120}, 243002 (2018).

\bibitem{Drut2} W.~S. Daza, J.~E. Drut, C.~L. Lin and C.~R. Ord{\'o}{\~n}ez, e-print arXiv:1808.0711v1 .

\bibitem{Pricoupenko3} A. Pricoupenko and D.~S. Petrov, Phys. Rev. A {\bf 97}, 063616 (2018).

\bibitem{Randeria} E. Taylor and M. Randeria, Phys. Rev. Lett. {\bf 109}, 135301 (2012).

\bibitem{Hofmann} J. Hofmann, Phys. Rev. Lett. {\bf 108}, 185303 (2012).
  
\bibitem{Beane2D} S.~R. Beane, Phys. Rev. A {\bf 82}, 063610 (2010).

\bibitem{LianyiHe} H. Hu, B.~C. Mulkerin, U. Toniolo, L. He and X.~-J. Liu, e-print arXiv:1806.04382 .

\bibitem{Hadzibabic} C. Eigen, J.~A.~P. Glidden, R. Lopes, N. Navon, Z. Hadzibabic and R.~P. Smith, Phys. Rev. Lett. {\bf 119}, 250404 (2017).

\bibitem{Pastukhov} V. Pastukhov, Phys. Lett. A (in press), https://doi.org/10.1016/j.physleta.2018.12.006 (2018).

\bibitem{Dalfovo} F. Dalfovo, Stefano Giorgini, L.~P. Pitaevskii and S. Stringari, Rev. Mod. Phys. {\bf 71}, 463 (1999).

\bibitem{Abramowitz} M. Abramowitz and I. Stegun, {\it Handbook of Mathematical Functions with Formulas, Graphs, and Mathematical Tables} (United States Department of Commerce, National Bureau of Standards 1964). 

\bibitem{Busch} T. Busch, B.~-G. Englert, K. Rzazewski and M. Wilkens, Found. Phys. {\bf 28}, 549 (1998).
  
\bibitem{Tan1} S. Tan, Ann. Phys. {\bf 323}, 2952 (2008).

\bibitem{Tan2} S. Tan, Ann. Phys. {\bf 323}, 2971 (2008).

\bibitem{Tan3} S. Tan, Ann. Phys. {\bf 323}, 2987 (2008).

\bibitem{Werner} F. Werner, Phys. Rev. A {\bf 78}, 025601 (2008).

\bibitem{ValienteZinnerMolmer1} M. Valiente, N.~T. Zinner and K. M{\o}lmer, Phys. Rev. A {\bf 84}, 063626 (2011). 

\bibitem{ValienteZinnerMolmer2}M. Valiente, N.~T. Zinner and K. M{\o}lmer, Phys. Rev. A {\bf 86}, 043616 (2012).
  
\bibitem{Matveev} Z. Ristivojevic and K.~A. Matveev, Phys. Rev. B {\bf 89}, 180507 (2014).

\bibitem{Lambert} R.~M. Corless, G.~H. Gonnet, D.~E.~G. Hare, D.~J. Jeffrey and D.~E. Knut, Adv. Comp. Math. {\bf 5}, 329 (1996).
  
\end{thebibliography}

\end{document}